\documentclass[apjl]{emulateapj}
\usepackage{psfig,amsfonts,amsmath,graphicx,apjfonts}
\usepackage{natbib}
\def\ra#1#2#3{#1$^{\rm h}$#2$^{\rm m}$#3$^{\rm s}$}
\def\dec#1#2#3{$#1^\circ#2'#3''$}

\def\grb{GRB\,090423}

\def\cfa{1}
\def\ssc{2}
\def\lei{3}
\def\war{4}
\def\col{5}

\begin{document}

\title{ALMA Observations of the Host Galaxy of GRB\,090423 at
  $z=8.23$: Deep Limits on Obscured Star Formation 630 Million Years
  After the Big Bang}

\author{
E.~Berger\altaffilmark{\cfa},
B.~A.~Zauderer\altaffilmark{\cfa},
R.-R.~Chary\altaffilmark{\cfa,}\altaffilmark{\ssc},
T.~Laskar\altaffilmark{\cfa},
R.~Chornock\altaffilmark{\cfa},
N.~R.~Tanvir\altaffilmark{\lei},
E.~R.~Stanway\altaffilmark{\war},
A.~J.~Levan\altaffilmark{\war},
E.~M.~Levesque\altaffilmark{\col}, \&
J.~E.~Davies\altaffilmark{\cfa}
}

\altaffiltext{1}{Harvard-Smithsonian Center for Astrophysics, 60
Garden Street, Cambridge, MA 02138, USA}

\altaffiltext{2}{Spitzer Science Center, California Institute of
  Technology, MC 220-6, 1200 East California Boulevard, Pasadena, CA
  91125, USA}

\altaffiltext{3}{Department of Physics and Astronomy, University of
  Leicester, University Road, Leicester LE1 7RH, UK}

\altaffiltext{4}{Department of Physics, University of Warwick, Gibbet
  Hill Road, Coventry CV4 7AL, UK}

\altaffiltext{5}{CASA, University of Colorado UCB 389, Boulder, CO
  80309, USA}

\begin{abstract} We present rest-frame far-infrared (FIR) and optical
  observations of the host galaxy of GRB\,090423 at $z=8.23$ from the
  Atacama Large Millimeter Array (ALMA) and the {\it Spitzer Space
    Telescope}, respectively.  The host remains undetected to
  $3\sigma$ limits of $F_\nu(222\,{\rm GHz})\lesssim 33$ $\mu$Jy and
  $F_\nu(3.6\,{\rm \mu m})\lesssim 81$ nJy.  The FIR limit is about 20
  times fainter than the luminosity of the local ULIRG Arp\,220, and
  comparable to the local starburst M\,82.  Comparing to model
  spectral energy distributions we place a limit on the IR luminosity
  of $L_{\rm IR}(8-1000{\rm \,\mu m})\lesssim 3\times 10^{10}$
  L$_\odot$, corresponding to a limit on the obscured star formation
  rate of ${\rm SFR_{\rm IR}} \lesssim 5$ M$_\odot$ yr$^{-1}$; for
  comparison, the limit on the unobscured star formation rate from
  {\it Hubble Space Telescope} rest-frame UV observations is ${\rm
    SFR_{UV}}\lesssim 1$ M$_\odot$ yr$^{-1}$.  We also place a limit
  on the host galaxy stellar mass of $M_*\lesssim 5\times 10^7$
  M$_\odot$ (for a stellar population age of 100 Myr and constant star
  formation rate).  Finally, we compare our millimeter observations to
  those of field galaxies at $z\gtrsim 4$ (Lyman break galaxies,
  Ly$\alpha$ emitters, and submillimeter galaxies), and find that our
  limit on the FIR luminosity is the most constraining to date,
  although the field galaxies have much larger rest-frame UV/optical
  luminosities than the host of GRB\,090423 by virtue of their
  selection techniques.  We conclude that GRB host galaxies at
  $z\gtrsim 4$, especially those with measured interstellar medium
  metallicities from afterglow spectroscopy, are an attractive sample
  for future ALMA studies of high redshift obscured star formation.
  \end{abstract}
 

\section{Introduction}
\label{sec:intro}

Thanks to their extreme brightness, gamma-ray bursts (GRBs) can be
utilized as powerful probes of star and galaxy formation at high
redshift, including during the epoch of reionization (e.g.,
\citealt{tkk+06,tfl+09,tlf+12,cbf+13,cbf+14}).  In particular,
spectroscopy of the optical/near-IR afterglow emission can provide an
unambiguous and precise spectroscopic redshift, as well as a
measurement of the host galaxy interstellar medium (ISM) properties
{\it independent of the host galaxy brightness} (e.g.,
\citealt{tkk+06,cbf+13}).  This unique insight can then be combined
with multi-wavelength photometric follow-up of the host galaxies to
study properties such as their stellar mass and star formation
activity (e.g., \citealt{lbc11,tlf+12}).

In contrast, direct galaxy studies at these redshifts rely either on
photometric redshifts (the Ly$\alpha$ break technique; LBGs), and are
therefore limited to bright objects usually with no subsequent
spectroscopic confirmation (e.g.,
\citealt{sbm03,bse+04,mrp+05,bib+06,oms+09,bio+11,mdd+11,vpf+11,oom+12,fpd+13}),
or on narrow-band Ly$\alpha$ imaging (Ly$\alpha$ emitters; LAEs),
which also requires subsequent spectroscopic confirmation (e.g.,
\citealt{ktk+03,iok+06,hcb+10,osf+10,sko+12}).  In the case of LAEs,
the reliance on Ly$\alpha$ emission may also inhibit detections during
the epoch of reionization due to significant scattering and absorption
of Ly$\alpha$ photons by the neutral intergalactic medium; the same
effect also inhibits the success of LBG spectroscopic confirmation,
which also relies on detection of the Ly$\alpha$ line.  Similarly, the
reliance on rest-frame UV and Ly$\alpha$ emission bias both the LBG
and LAE selection against dusty environments.  Moreover, for both LBGs
and LAEs, measurements of the ISM properties are beyond the reach of
current instrumentation since the relevant rest-frame optical emission
lines are redshifted into the mid-IR.  An alternative technique of
submillimeter color selection has also led to the identification of a
few high redshift submillimeter-bright galaxies (SMGs), with
spectroscopic confirmation via CO and [\ion{C}{2}] emission lines
(e.g., \citealt{wdc+12,rbc+13}).  However, these extremely luminous
and highly star forming galaxies are exceedingly rare and do not shed
light on the bulk of the galaxy population.  Thus,
spectroscopically-confirmed GRB host galaxies at high redshift provide
an attractive sample for multi-wavelength follow-up observations.

In this context, the advent of the Atacama Large Millimeter Array
(ALMA) makes it possible for the first time to probe obscured star
formation in high-redshift galaxies to a sensitivity level of only a
few M$_\odot$ yr$^{-1}$, or an integrated IR luminosity of $L_{\rm
  IR}\sim 10^{10}$ L$_\odot$.  Previously, only rare SMGs with $L_{\rm
  IR}\gtrsim 10^{12}$ L$_\odot$ (ultra-luminous infrared galaxies;
ULIRGs) and correspondingly high star formation rates of $\sim 10^3$
M$_\odot$ yr$^{-1}$ were accessible (e.g.,
\citealt{ckn+11,crr+12,wdc+12,rbc+13}).  In the long term, ALMA
observations of GRB host galaxies with measured ISM metallicities will
shed light on the relations between metallicity, dust content, and
obscured star formation at redshifts that are inaccessible with other
techniques (i.e., $z\gtrsim 4$).  This will also provide a comparison
with the rest-frame FIR properties of LBGs and LAEs, although none
have been individually detected to date at $z\gtrsim 4$ (e.g.,
\citealt{oeo+13,owo+14}).

Here, we present deep ALMA observations of the host galaxy of the most
distant spectroscopically-confirmed burst to date -- \grb\ at $z=8.23$
\citep{sdc+09,tfl+09}.  The observations are aimed at reaching a
rest-frame FIR luminosity level below that of a luminous infrared
galaxy (LIRG: $L_{IR}=10^{11}$ L$_\odot$).  We also present {\it
  Spitzer Space Telescope} rest-frame optical observations to
constrain the stellar mass of the host galaxy.  The ALMA and {\it
  Spitzer} observations and data analysis are described in
\S\ref{sec:obs}.  In \S\ref{sec:prop} we use the data to constrain the
host galaxy properties, while in \S\ref{sec:comp} we compare these
results to millimeter studies of LAEs, LBGs, and SMGs at $z\gtrsim 4$.

\section{Observations} 
\label{sec:obs}

The discovery and redshift determination of \grb\ are presented in
\citet{tfl+09}, who find $z=8.23^{+0.06}_{-0.07}$.  The position of
the afterglow is R.A.\,=\,\ra{09}{55}{33.29},
Decl.\,=\,\dec{+18}{08}{57.8} (J2000) with an uncertainty of about
$0''.1$ in each coordinate.  The host galaxy was previously observed
with the {\it Hubble Space Telescope} Wide-field Camera 3 ({\it
  HST}/WFC3) in the F110W and F160W filters, leading to non-detections
\citep{tlf+12}.  It was also observed with the Australia Telescope
Compact Array (ATCA), yielding non-detections of rest-frame 850 $\mu$m
continuum emission and CO (3-2) emission \citep{sbt+11}.  The
continuum limits are summarized in Table~\ref{tab:data}.

\subsection{Atacama Large Millimeter Array}

ALMA band 6 (222 GHz) observations were carried out with $25-28$ 12-m
diameter antennas in five observing blocks spread over 12 days on 2013
November 18, 21, and 30 UT (Cycle 1).  The 2 GHz wide spectral windows
were set to central frequencies of 213, 215, 229 and 231 GHz.  The
total on-source integration time was 163 min after data flagging.  We
performed data calibration and imaging using the Common Astronomy
Software Application (CASA).  The gain calibration utilized
J\,0854+2006, while bandpass calibration utilized one of
J\,0522$-$3627, J\,0538$-$4405, J\,1037$-$2934, or J\,1058+0133 in
each observing block.  Absolute flux calibration was performed using
observations of the solar system objects Ceres, Ganymede, or Pallas.
We found some minor variations in the resulting fluxes of J\,0854+2006
between the observing blocks, likely due to the use of asteroids,
which have uncertain flux models.  We therefore used Ganymede when
possible and Ceres to determine and manually set the flux for
J\,0854+2006 in each epoch\footnotemark\footnotetext{We used the
  following values for each spectral window and date: Nov.~18: 2.43,
  2.39, 2.29, and 2.28 Jy; Nov.~21: 2.25, 2.24, 2.17, and 2.16 Jy; and
  Nov.~30: 1.81, 1.80, 1.74, and 1.74 Jy.}.  The overall uncertainty
in the absolute flux calibration is about $10\%$.

\begin{figure}
\begin{center}
\includegraphics[angle=0,width=6.5in]{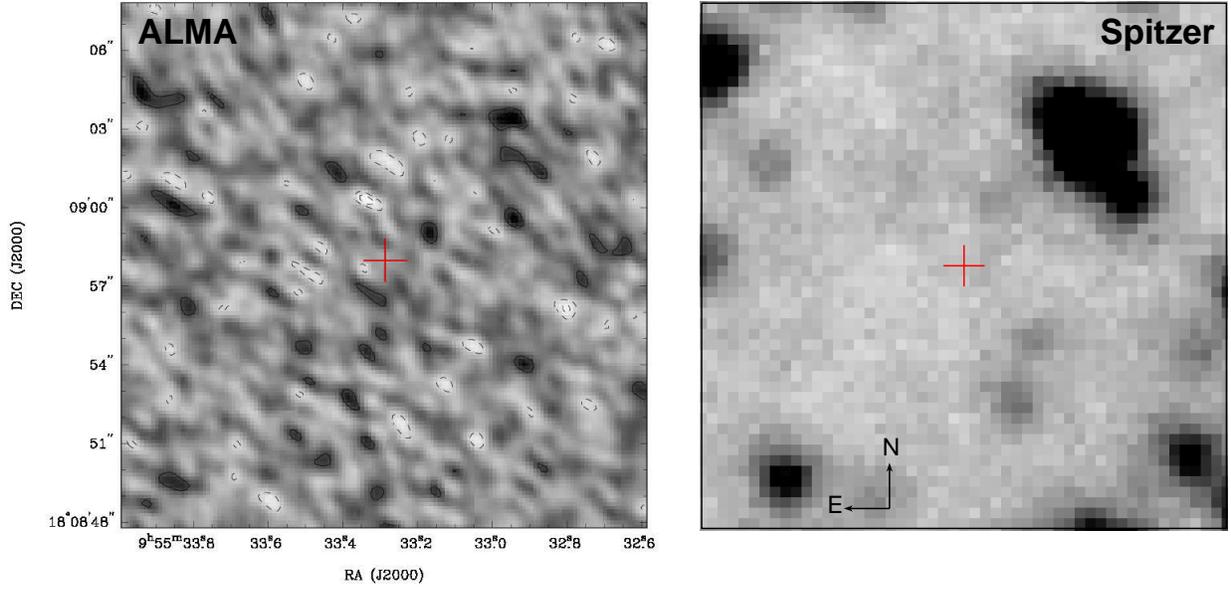}
\end{center}
\caption{{\it Left:} ALMA band 6 continuum image of a $20''\times
  20''$ region centered on the location of the host galaxy of \grb\
  (cross).  Contours are in steps of $1\sigma=11$ $\mu$Jy beam$^{-1}$
  starting at $\pm 2\sigma$ (solid: positive; dashed: negative).  No
  millimeter emission is detected at the location of the host galaxy.
  {\it Right:} {\it Spitzer}/IRAC 3.6 $\mu$m image of a $20''\times
  20''$ region centered on the location of the host galaxy of \grb\
  (cross).  No infrared emission is detected at the location of the
  host galaxy. 
  \label{fig:image}}
\end{figure}

The resulting combined continuum map (natural weighting without
primary beam correction) has an rms noise of $11$ $\mu$Jy beam$^{-1}$
for a synthesized beam size (full-width at half maximum) of
$1''.0\times 0''.80$ (Figure~\ref{fig:image}).  No source is detected
at the location of \grb\footnotemark\footnotetext{We note the presence
  of a $3\sigma$ peak at R.A.~=\ra{09}{55}{33.17},
  Decl.~=\dec{+18}{08}{59.0} (J2000), about $2''$ away from the
  position of \grb, but given the low signal-to-noise ratio and the
  substantial offset we do not consider this potential source to be
  related to \grb.} (with a flux density at the source location of
$-4\pm 11$ $\mu$Jy), leading an upper limit of $F_\nu(222\,{\rm
  GHz})\lesssim 33$ $\mu$Jy ($3\sigma$; Table~\ref{tab:data}).

\subsection{{\it Spitzer Space Telescope}}

{\it Spitzer} observations were obtained during the warm mission (DDT
Program 538) with the Infrared Array Camera (IRAC) at $3.6$ $\mu$m on
2010 January 26 and 27 UT.  Each observation consisted of 650 100-s
frames ($93.6$ s on source), for a combined on-source time of $33.8$
hr.  We analyzed the data using the Great Observatories Origins Deep
Survey (GOODS) pipeline, including corrections for bright star image
artifacts (muxbleed and pulldown) with dark and sky background
subtraction.  We carried out astrometric alignment utilizing stars in
common with the Sloan Digital Survey Survey database, and created a
combined mosaic of the individual frames on a $0.4''\times 0.4''$
pixel grid.  We also produced a pair of split images, each with half
of the frames, and used the difference in these split images to
produce a noise map.  The resulting uncertainty is about 2 times
higher than that derived from pure pixel statistics in the vicinity of
the source.

We do not detect significant emission at the location of \grb\
(Figure~\ref{fig:image}).  There is no evidence for significant source
confusion at this location, in agreement with the {\it HST}/WFC3
images of the field \citep{tlf+12}.  From the noise maps, we find that
the $1\sigma$ point source sensitivity at the location of \grb\ is 27
nJy.  Photometry in a $1.2"$ radius aperture yields a $3.6$ $\mu$m
flux density of $26\pm 27$ nJy, after aperture corrections.  We
therefore place an upper limit on the brightness of the host of
$F_\nu(3.6\,{\rm \mu m}\lesssim 81$nJy ($3\sigma$;
Table~\ref{tab:data}).

\section{Host Galaxy Properties: IR Luminosity, Obscured Star
  Formation Rate, and Stellar Mass}
\label{sec:prop}

\begin{figure}
\begin{center}
\includegraphics[angle=0,height=2.75in]{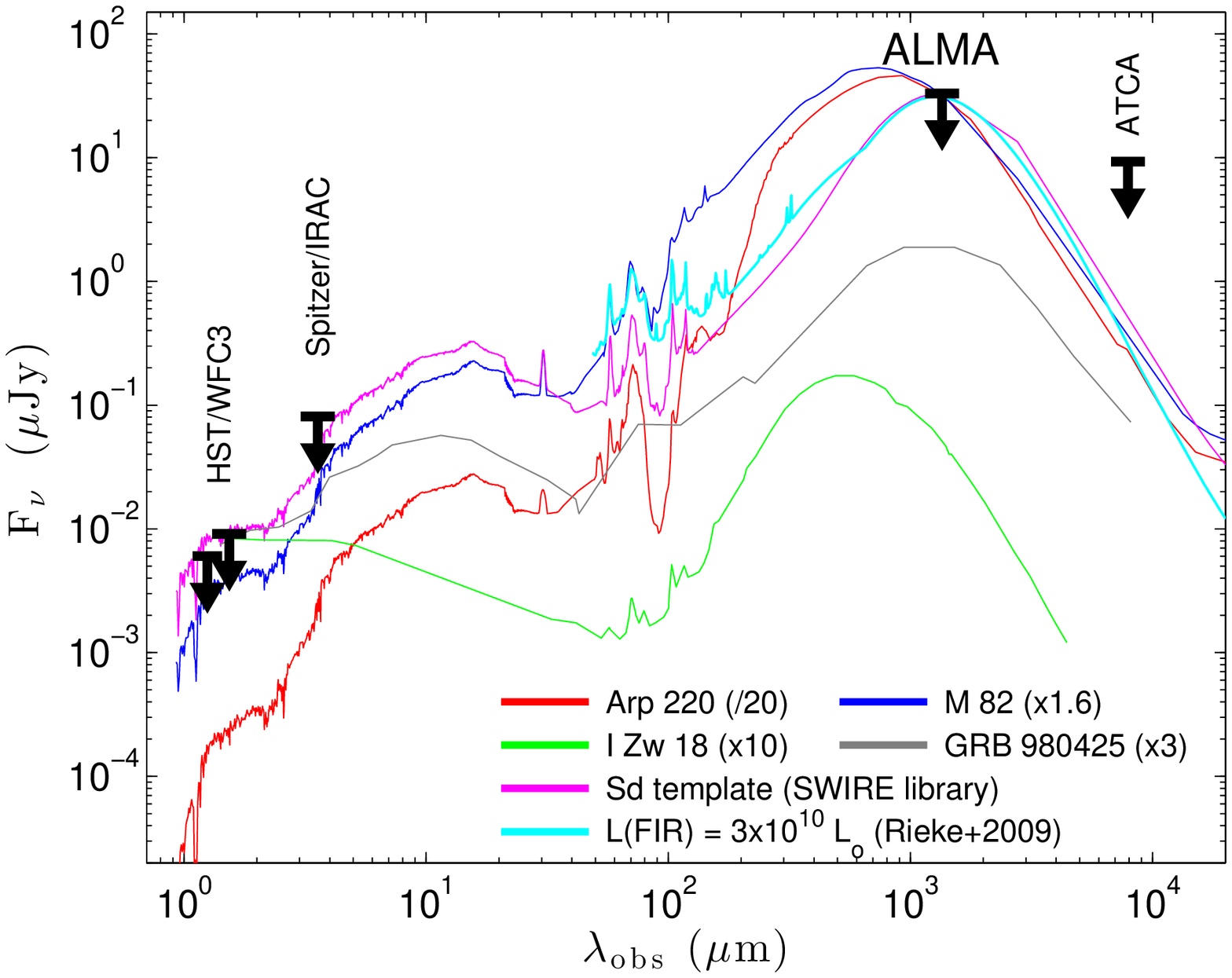}\hfill
\includegraphics[angle=0,height=2.75in]{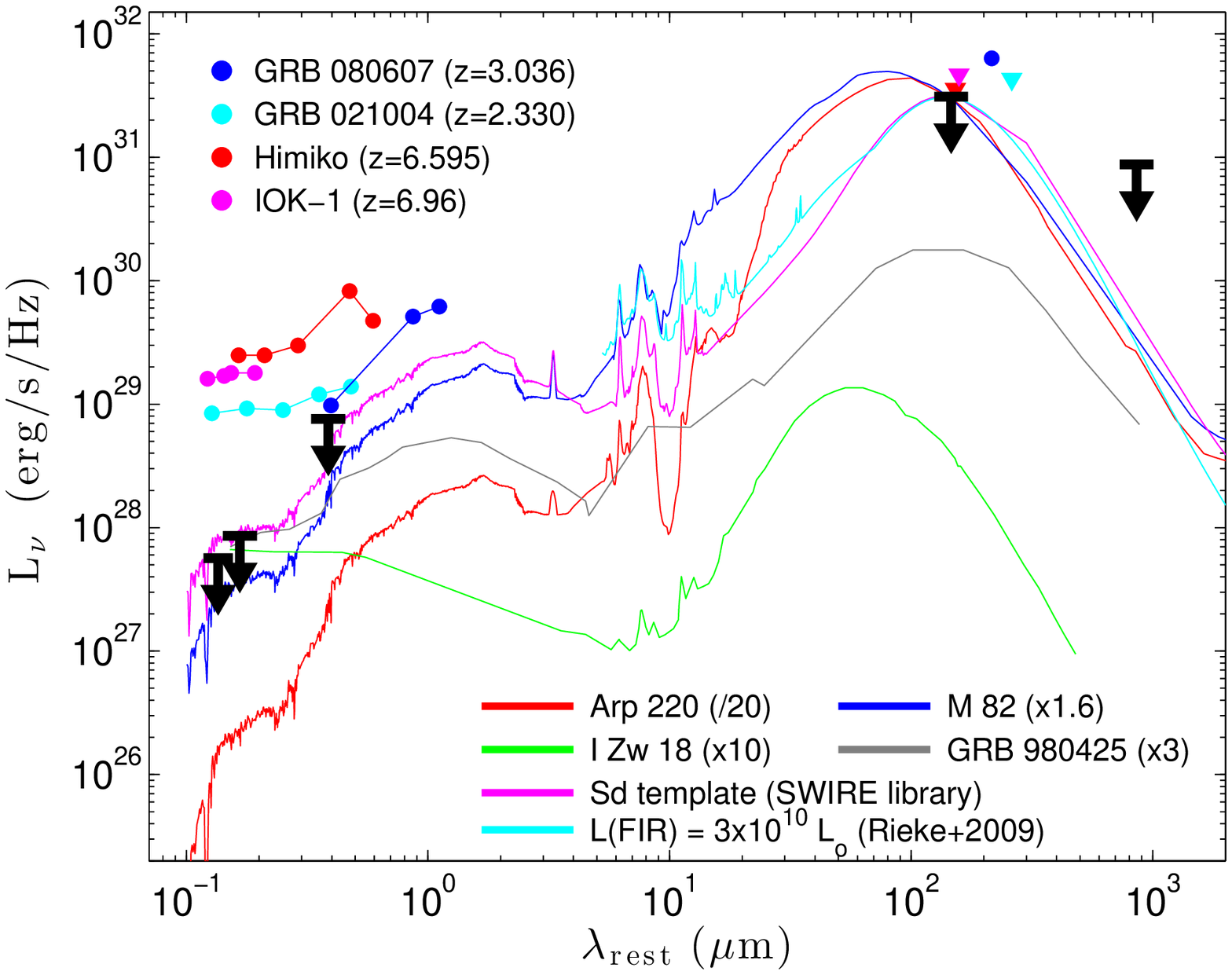}
\end{center}
\caption{{\it Left:} Limits on the flux density of the host galaxy of
  \grb\ in the near-IR ({\it HST}), mid-IR ({\it Spitzer}), millimeter
  (ALMA), and radio (ATCA).  Also shown are the SEDs of the local
  ULIRG Arp\,220 (red), the local starburst M\,82 (blue), the local
  dwarf I\,Zw\,18 (green), the local host galaxy of GRB\,980425 (gray;
  \citealt{mhp+14}), an Sd galaxy template (magenta), and a template
  for a galaxy with $L_{\rm IR}=3\times 10^{10}$ L$_\odot$ (cyan;
  \citealt{raw+09}) all shifted to $z=8.23$; with the exception of
  I\,Zw\,18 and the host of GRB\,980425, which are scaled to the {\it
    HST} limits, the galaxy models are scaled to match the ALMA flux
  density limit.  For the Arp\,220 template the ALMA non-detection
  places a stronger constraint on the SED than the {\it HST} and {\it
    Spitzer} limits, while for the starburst templates the limits are
  comparable.  For the I\,Zw\,18 and GRB\,980425 host galaxy
  templates, the {\it HST} limits are more constraining.  {\it Right:}
  Same as the left panel, but plotting the rest-frame luminosity
  density and wavelength.  Also shown are the ALMA observations and
  rest-frame UV/optical SEDs of two other GRB host galaxies (circles: detections;
  triangles: upper limits; GRB\,080607 is a marginal $3.4\sigma$
  detection; \citealt{wch+12}), and two spectroscopically-confirmed
  LAEs at $z\approx 6.6-7.0$ \citep{oeo+13,owo+14}. 
  \label{fig:seds}}
\end{figure}

The host galaxy of \grb\ remains undetected in our ALMA and {\it
  Spitzer} observations.  The flux density limits are plotted in the
left panel of Figure~\ref{fig:seds}.  Converted to rest-frame spectral
luminosity, the non-detections correspond to $L_\nu(145\,{\rm \mu
  m})\lesssim 3.1\times 10^{31}$ erg s$^{-1}$ Hz$^{-1}$ and
$L_\nu(0.39\,{\rm \mu m})\lesssim 7.6\times 10^{28}$ erg s$^{-1}$
Hz$^{-1}$, respectively (right panel of Figure~\ref{fig:seds}, and
Table~\ref{tab:data}).

Also shown in Figure~\ref{fig:seds} are the spectral energy
distributions (SEDs) of several local galaxies redshifted to $z=8.23$:
the ULIRG Arp\,220, the starburst M\,82, the low metallicity dwarf
galaxy I\,Zw\,18, and the host galaxy of GRB\,980425 ($d=40$ Mpc;
\citealt{mhp+14}).  Our ALMA non-detection corresponds to a luminosity
that is about 20 times fainter than that of Arp\,220 and about 1.6
times higher than that of M\,82.  We also provide a comparison with
the galaxy templates of \citet{raw+09}, which are based on eleven
local star forming LIRGs and ULIRGs, and find that the ALMA
non-detection corresponds to $L_{\rm IR}\lesssim 3\times 10^{10}$
L$_\odot$.

For the Arp\,220 template, the ALMA non-detection provides a much more
stringent limit than the {\rm HST} and {\it Spitzer} limits, which
only rule out a luminosity comparable to that of Arp\,220.  The ALMA
constraints are also more stringent for the M\,82 template, although
only by about a factor of 2.  For an Sd galaxy template from the
Spitzer Wide-area InfraRed Extragalactic survey (SWIRE) library, the
ALMA, {\it HST}, and {\it Spitzer} data all place comparable limits on
the host SED.  On the other hand, for the I\,Zw\,18 template, which
has weak FIR emission, the {\it HST} limits place a much more
stringent constraint than the ALMA data, by about a factor of 600.
The same is true for the GRB\,980425 host galaxy template, which is
about 15 times fainter than the ALMA limit when scaled to the {\it
  HST} limits.

Using the standard modified blackbody SED of dust emission, with a
dust temperature range of $T_{\rm dust}\approx 30-50$ K and
$\beta\approx 1.5-2$ (e.g., \citealt{sss+14}), the ALMA non-detection
corresponds to an upper limit on the integrated IR luminosity of
$L_{\rm IR}(8-1000\, {\rm \mu m})\lesssim (2-5)\times 10^{10}$
L$_\odot$; the lower bound on the blackbody temperature is set by the
cosmic microwave background temperature $T_{\rm CMB}(z=8.23)\approx
25$ K \citep{dgw+13}.  The IR luminosity upper limit is thus a few
times lower than the scale for a LIRG, and agrees well with the
comparison to the \citet{raw+09} templates.  The limit on the
integrated IR luminosity corresponds to an upper bound on the obscured
star formation rate \citep{ken98,cwh+10},
of\footnotemark\footnotetext{Using the scaled SED of I\,Zw\,18, which
  has much weaker FIR emission relative to its optical/UV emission,
  leads to a limit of ${\rm SFR}_{\rm IR}\lesssim 100$ M$_\odot$
  yr$^{-1}$.}  ${\rm SFR}_{\rm IR}\lesssim 3-5$ M$_\odot$ yr$^{-1}$.
For comparison, the limit on the {\it unobscured} star formation rate
from the {\it HST} rest-frame UV non-detection is ${\rm SFR}_{\rm UV}
\lesssim 1.2$ M$_\odot$ yr$^{-1}$.

We also place an upper bound on the host galaxy stellar mass using the
{\it HST} and {\it Spitzer} upper limits.  Utilizing the \citet{bc03}
stellar population synthesis models with a constant star formation
rate, a Salpeter IMF, a metallicity of $0.2$ Z$_\odot$, and no dust
extinction, we find that for a stellar population age of $10-100$ Myr
(typical of GRB host galaxies; \citealt{sgl09,lb10}) the limit on the
host stellar mass is $M_*\lesssim (1-5)\times 10^7$ M$_\odot$.  The
inferred host galaxy properties are summarized in
Table~\ref{tab:host}.

\section{Comparison to Millimeter Observations of Field Galaxies at
  $z\gtrsim 4$}
\label{sec:comp}

\begin{figure}
\begin{center}
\includegraphics[angle=0,height=4.5in]{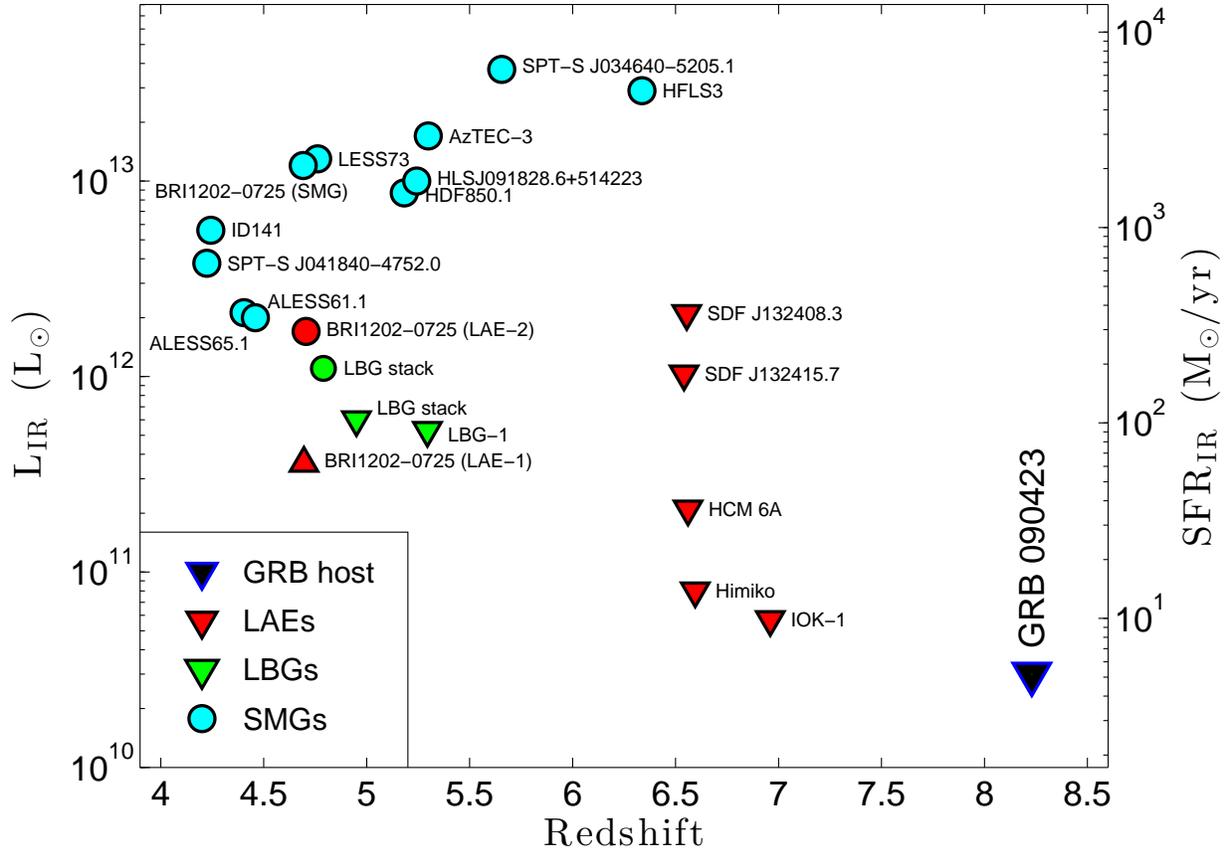}
\end{center}
\caption{Integrated IR luminosities and inferred obscured star
  formation rates (${\rm SFR_{IR}}=4.5\times 10^{-44}\,L_{\rm IR}$
  M$_\odot$ yr$^{-1}$; \citealt{ken98}) of galaxies with spectroscopic
  redshifts of $z\gtrsim 4$ observed with ALMA and other millimeter
  telescopes (circles: detections; triangles: upper limits).  The
  objects include SMGs (cyan), LBGs (green), LAEs (red), and the host
  of \grb\ (black).  So far, mostly SMGs have been detected as
  individual galaxies at $z\gtrsim 4$.  See \S\ref{sec:comp} for
  details and references. 
  \label{fig:gals}}
\end{figure}

To place the non-detection of the host galaxy of \grb\ in the context
of millimeter studies of high-redshift galaxies, we summarize below
previous observational efforts to detect and study distant LAEs, LBGs,
and SMGs; the information is summarized in Figure~\ref{fig:gals}.  We
first note that, at present, ALMA observations of only two other
galaxies with spectroscopic redshifts of $z\gtrsim 6$ have been
published: the LAEs ``Himiko'' at $z=6.595$ \citep{oeo+13,owo+14} and
IOK-1 at $z=6.96$ \citep{owo+14}.  Neither source was detected, with
resulting $3\sigma$ flux density limits of $F_\nu(259\,{\rm GHz})
\lesssim 52$ and $F_\nu(232\,{\rm GHz})\lesssim 63$ $\mu$Jy,
respectively.  The flux density and integrated IR luminosity limits
for these two LAEs are shallower than those achieved in our
observations; see Figure~\ref{fig:gals}.  However, it is important to
note that Himiko and IOK-1 are significantly more luminous than the
host of \grb\ in the rest-frame UV and optical, by a factor of
$\gtrsim 30$ (Figure~\ref{fig:seds}), with inferred unobscured star
formation rates of $\approx 25-30$ M$_\odot$ yr$^{-1}$
\citep{oeo+13,owo+14}.  Thus, in terms of the ratio of FIR to UV
luminosity, the limits for Himiko and IOK-1 are more constraining than
those for the host of \grb.

Beyond the ALMA observations of $z\gtrsim 6$ galaxies, there are
shallower observations from other millimeter
facilities. \citet{bsp+07} observed the lensed LAE HCM\,6A at $z=6.56$
with the Plateau de Bure Interferometer and placed a limit of $L_{\rm
  IR}\lesssim 2.1\times 10^{11}$ L$_\odot$.  \citet{grd+14} used the
Combined Array for Research in Millimeter-wave Astronomy to place
limits on two LAEs at $z=6.541$ and $z=6.554$ of $\lesssim 1.0\times
10^{12}$ and $2.1\times 10^{12}$ L$_\odot$, respectively.
\citet{rbc+13} used a wide range of facilities to study the bright SMG
HFLS\,3 at $z=6.337$, discovered in the Herschel Multi-tiered
Extragalactic Survey, and found $L_{\rm IR} \approx 2.9\times 10^{13}$
L$_\odot$.  Thus, the only galaxy at $z\gtrsim 6$ detected in the
millimeter band to date is a hyper-luminous infrared galaxy
(Figure~\ref{fig:gals}).

Expanding the redshift range down to $z=4$, only a few galaxies with
spectroscopic redshifts have been detected with ALMA, all of which are
bright SMGs (Figure~\ref{fig:gals}).  \citet{sks+12} detected two SMGs
in ALMA observations of the LABOCA Extended Chandra Deep Field-South
Survey, with [\ion{C}{2}]$\lambda 157.74$ $\mu$m detections indicating
redshifts of $z=4.419$ and $z=4.444$, and resulting IR luminosities of
$2\times 10^{12}$ L$_\odot$.  \citet{wwc+12} and \citet{crw+13}
presented continuum and [\ion{C}{2}] line emission from the SMG and
two LAEs associated with the AGN/SMG system BRI\,1202-0725 at
$z=4.70$, with luminosities of $1.2\times 10^{13}$ L$_\odot$ (SMG),
$\gtrsim 3.6\times 10^{11}$ L$_\odot$ (Ly$\alpha$-1), and $1.7\times
10^{12}$ L$_\odot$ (Ly$\alpha$-2).  \citet{hmf+13} detected two
strongly lensed SMGs at $z=4.224$ and $z=5.656$, found in South Pole
Telescope data, with intrinsic IR luminosities of about $3.8\times
10^{12}$ and $3.7\times 10^{13}$ L$_\odot$, respectively.
\citet{wcd+14} detected an SMG at $z=4.762$, first identified in
LABOCA observations of the Extended Chandra Deep Field, with $L_{\rm
  IR}\approx 1.3\times 10^{13}$.  \citet{rcc+14} detected the SMG
AzTEC-3 at $z=5.299$ with $L_{\rm IR}\approx 1.7\times 10^{13}$
L$_\odot$, and placed a limit on an LBG in the same field at a similar
redshift of $z=5.295$ of $L_{\rm IR}\lesssim 5.3 \times 10^{11}$
L$_\odot$.

In addition to the ALMA observations, \citet{ckn+11} studied ID\,141,
a lensed galaxy at $z=4.243$ from the Herschel Astrophysical Terahertz
Large Area Survey, and inferred $L_{\rm IR}\approx (3-8)\times
10^{12}$ L$_\odot$; the range of values accounts for the unknown
lensing magnification factor.  \citet{crr+12} discovered and studied a
lensed SMG at $z=5.243$ from the Herschel Lensing Survey with $L_{\rm
  IR}\approx 10^{13}$ L$_\odot$.  \citet{wdc+12} determined a redshift
of $z=5.183$ for the SMG HDF\,850.1 based on CO and [\ion{C}{2}] line
emission, leading to $L_{\rm IR} \approx 8.7\times 10^{12}$ L$_\odot$.
\citet{dbs+12} used stacked MAMBO-2 observations of $z\sim 5$ LBGs to
place a limit on their mean luminosity of $L_{\rm IR}\lesssim 3\times
10^{11}$ L$_\odot$, while \citet{cga+14} used stacked SCUBA-2
observations of $z\sim 4.8$ LBGs to find a mean luminosity of $L_{\rm
  IR}\sim 10^{12}$ L$_\odot$ (based on a $3.8\sigma$ detection at 850
$\mu$m).

Finally, two other GRB host galaxies at lower redshifts have been
observed with ALMA \citep{wch+12}.  The host of GRB\,021004 at
$z=2.330$ was not detected, with a resulting limit of $L_{\rm IR}
\lesssim 3\times 10^{11}$ L$_\odot$ (${\rm SFR_{IR}}\lesssim 50$
M$_\odot$ yr$^{-1}$).  The host of the dusty GRB\,080607 at $z=3.036$
was marginally detected ($3.4\sigma$) with a resulting luminosity of
$(2.4-4.5)\times 10^{11}$ L$_\odot$ (${\rm SFR_{IR}}\approx 40-80$
M$_\odot$ yr$^{-1}$).  Prior to ALMA, a few GRB host galaxies at
$z\lesssim 2$ were detected in the submillimeter and radio bands
\citep{bkf01,fbm+02,bck+03,mhp+14,sod+14}, some with inferred IR
luminosities that exceed that of Arp\,220.  However, most GRB host
galaxies were not detected at a comparable level to Arp\,220 or
fainter \citep{bck+03,tbb+04,ptl+06,pp13}, ruling out a dominant ULIRG
host population.

Thus, in comparison to the previous studies of $z\gtrsim 4$
spectroscopically-confirmed field galaxies, as well as other GRB host
galaxies, our ALMA observations of the host of \grb\ represent the
deepest limit to date in terms of flux density, spectral luminosity,
and integrated IR luminosity (Figure~\ref{fig:gals}).

\section{Conclusions}
\label{sec:conc}

We present ALMA and {\it Spitzer} observations of the host galaxy of
\grb\ at $z=8.23$, the highest redshift spectroscopically-confirmed
galaxy observed with these facilities to date.  The host galaxy
remains undetected at rest-frame wavelength of $145$ $\mu$m (ALMA) and
$0.39$ $\mu$m ({\it Spitzer}).  The resulting limit on the integrated
IR luminosity is $L_{\rm IR}\lesssim 3\times 10^{10}$ L$_\odot$,
corresponding to $\lesssim 5$ M$_\odot$ yr$^{-1}$ of obscured star
formation; scaling the SED of the low metallicity dwarf galaxy
I\,Zw\,18 relaxes the star formation rate upper limit to $\lesssim
100$ M$_\odot$ yr$^{-1}$.  In addition, based on the {\it Spitzer} and
{\it HST} non-detections we place a limit on the host galaxy stellar
mass of $M_*\lesssim 5\times 10^7$ M$_\odot$ (100 Myr old stellar
population with constant star formation rate).  The limit on the
unobscured star formation rate based on {\it HST} rest-frame UV
observations is $\lesssim 1.2$ M$_\odot$ yr$^{-1}$.

We additionally compare our ALMA non-detection to millimeter
observations of spectroscopically-confirmed galaxies at $z\gtrsim 4$
(undertaken with ALMA and other facilities) and show that the limit on
the host of \grb\ is the deepest to date compared to any published
observations.  At this redshift range only SMGs have been convincingly
detected, while individual LBGs and LAEs have so far escaped detection
even with ALMA (a marginal detection of LBGs at $z\sim 4.8$ in stacked
SCUBA-2 observations has been reported by \citealt{cga+14}).  The only
comparable limits to ours are based on ALMA observations of two LAEs
at $z=6.595$ and $z=6.96$.

It is quite remarkable that we were able to reach a brightness limit
of about 20 times lower than Arp\,220 (and comparable to M\,82) at
$z=8.23$ in only 3 hours of on-source time and with a subset of the
ALMA antennas.  On the other hand, if typical high redshift galaxies
have SEDs similar to the local dwarf I\,Zw\,18 or the host galaxy of
GRB\,980425, then much deeper ALMA observations will be required to
detect their FIR emission.  Still, looking forward we anticipate that
ALMA observations of GRB host galaxies will be highly desirable and
productive, especially targeting spectroscopically-confirmed hosts at
$z\gtrsim 4$ with detailed ISM metallicity measurements, a unique
sample among high-redshift galaxies.

\acknowledgements The Berger GRB group at Harvard is supported in part
by the National Science Foundation under Grant AST-1107973.  This
paper makes use of the following ALMA data:
ADS/JAO.ALMA\#2012.1.00953.S.  ALMA is a partnership of ESO
(representing its member states), NSF (USA) and NINS (Japan), together
with NRC (Canada) and NSC and ASIAA (Taiwan), in cooperation with the
Republic of Chile.  The Joint ALMA Observatory is operated by ESO,
AUI/NRAO and NAOJ.  The National Radio Astronomy Observatory is a
facility of the National Science Foundation operated under cooperative
agreement by Associated Universities, Inc.  This work is based in part
on observations made with the Spitzer Space Telescope, which is
operated by the Jet Propulsion Laboratory, California Institute of
Technology under a contract with NASA.

{\it Facilities:} \facility{ALMA}, \facility{Spitzer (IRAC)}

\clearpage
\begin{deluxetable}{lllll}
\tablecolumns{5}
\tabcolsep0.1in\footnotesize
\tablewidth{0pc}
\tablecaption{Observations of the Host Galaxy of \grb\
\label{tab:data}}
\tablehead {
\colhead{Instrument}              &
\colhead {$\lambda_{\rm obs}$} &
\colhead {$F_\nu\,^a$}             &
\colhead {$L_\nu\,^a$}             &
\colhead{Reference}                \\
\colhead {}                              &
\colhead {}                              &
\colhead {($\mu$Jy)}              &
\colhead {(erg s$^{-1}$ Hz$^{-1}$)} &                    
\colhead {}                              
}
\startdata
ATCA & 0.79 cm & $\lesssim 9.3$ & $\lesssim 8.7\times 10^{30}$ & Stanway et al.~2011 \\
ALMA & 0.14 cm & $\lesssim 33$ & $\lesssim 3.1\times 10^{31}$ & This paper \\
{\it Spitzer}/IRAC & 3.6 $\mu$m & $\lesssim 8.1\times 10^{-2}$ & $\lesssim 7.6\times 10^{28}$ & This paper \\
{\it HST}/WFC3    & 1.54 $\mu$m & $\lesssim 9.1\times 10^{-3}$ & $\lesssim 8.6\times 10^{27}$ & Tanvir et al.~2012 \\
{\it HST}/WFC3    & 1.25 $\mu$m & $\lesssim 6.0\times 10^{-3}$ & $\lesssim 5.6\times 10^{27}$ & Tanvir et al.~2012 
\enddata
\tablecomments{$^a$ Limits are $3\sigma$.}
\end{deluxetable}

\clearpage
\begin{deluxetable}{ll}
\tablecolumns{2}
\tabcolsep0.5in\footnotesize
\tablewidth{0pc}
\tablecaption{Inferred Properties of the Host Galaxy of \grb\
\label{tab:host}}
\tablehead {
\colhead{Parameter} &
\colhead {Value}           
}
\startdata
$L_{\rm IR}\,^a$ & $\lesssim (2-5)\times 10^{10}$ L$_\odot$ \\
${\rm SFR_{IR}}\,^a$ & $\lesssim 3-5$ M$_\odot$ yr$^{-1}$  \\
${\rm SFR_{UV}}\,^b$ & $\lesssim 1.2$ M$_\odot$ yr$^{-1}$  \\
$M_*\,^c$ & $\lesssim (1-5)\times 10^7$ M$_\odot$ 
\enddata
\tablecomments{$^a$ Assuming $T_{\rm dust}\approx 30-50$ K and
  $\beta\approx 1.5-2$. \\ $^b$ Assuming $A_V^{\rm host}=0$ mag. \\
  $^c$ Assuming a stellar population age of $10-100$ Myr, constant
  star formation rate, Salpeter IMF, $Z=0.2$ Z$_\odot$, and $A_V^{\rm
    host}=0$ mag.}
\end{deluxetable}

\clearpage

\clearpage

\clearpage

\end{document}